\documentclass[aps,prl,twocolumn]{revtex4}
\usepackage{epsfig}
\usepackage{graphics}
\usepackage{graphicx}

\begin{document}
\title{Indirect resonant inelastic X-ray scattering on magnons}
\author{Jeroen van den Brink}
\affiliation{
Institute-Lorentz for Theoretical Physics,  Universiteit  Leiden,\\
P.O. Box 9506, 2300 RA Leiden,The Netherlands and\\
 Institute for Molecules and Materials, Radboud Universiteit Nijmegen,\\ 
P.O. Box 9010, 6500 GL Nijmegen, The Netherlands
}
\date{\today}
\begin{abstract}
Recent experiments show that indirect resonant inelastic X-ray scattering (RIXS) is a new probe of spin dynamics. Here I derive the cross-section for magnetic RIXS and determine the momentum dependent four-spin correlation function that it measures. These results show that this technique offers information on spin dynamics that is complementary to e.g. neutron scattering. The RIXS spectrum of Heisenberg antiferromagnets is calculated. It turns out that only scattering processes that involve at least two magnons are allowed. Other selection rules imply that the scattering intensity vanishes for specific transferred momenta ${\bf q}$, in particular for ${\bf q}=0$. The calculated spectra agree very well with the experimental data. 
\end{abstract}
\maketitle
%\begin{multicols}{1}
%\narrowtext

{\it Introduction.} 
Resonant inelastic X-ray scattering (RIXS) is a technique that is rapidly developing due to the recent increase in brilliance of the new generation synchrotron X-ray sources~\cite{Kotani01}. As a scattering technique  RIXS has two important advantages. First, it is sensitive to excitations that are difficult to observe by otherwise, for example direct $d$-$d$ excitations in cuprates or manganites~\cite{Hasan00,Hasan02,Kim02,Hill98,Hamalainen00,Inami03}. Second, it probes directly both the energy and momentum dependence of such excitations --unlike for instance optical methods. 

Experiments are performed at the K-edges of transition metal ions, where the frequency of the incident hard X-rays is tuned to match the energy of an atomic $1s$-$4p$ transition (5-10 keV). When a photon is absorbed in the solid a $1s$ core-hole is created as the electron is promoted to the $4p$ empty states. From the theoretical side it is clear that the scattering intensity is related to the {\it charge dynamics} of the system under study~\cite{Isaacs96,Abbamonte99,Tsutsui03,Doering04}. The precise nature of this relation was established recently:  a systematic expansion in the ultra-short life time of the core-hole~\cite{Brink05,Ament06} shows that the effective RIXS cross section a resonant scattering factor times a combination of the linear charge response function $S({\bf q},\omega)$ and the dynamic {\it longitudinal spin} density correlation function of the $d-$electrons. 
%This result was shown to be asymptotically exact for both strong and weak local core-hole potentials. 

A recent breakthrough was achieved by J.P. Hill {\it et al.}~\cite{Hill05}, who observed that RIXS on the high temperature superconductors La$_{2-x}$Cu$_x$O$_4$ picks up {\it transversal spin dynamics} --magnons. This triggers the question what kind of transversal spin correlation function actually is measured in magnetic RIXS and how this is different from other probes of spin dynamics --in particular inelastic neutron scattering.  The aim of this Letter is to answer that question.

\begin{figure}
\includegraphics[width=0.9\columnwidth]{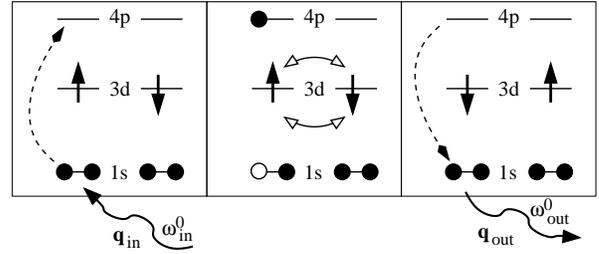} 
\caption{Schematic representation of the magnetic RIXS scattering process at a transition metal K-edge. Left: the incoming photon (energy $\omega^0_{\rm in}$, momentum ${\bf q}_{\rm in}$) induces an electronic transition from a $1s$ to a $4p$ level. Middle:  exchange interaction between $3d$ electrons in the presence of the core-hole. Right: de-excitation and outgoing photon ($\omega^0_{\rm out}$, ${\bf q}_{\rm out}$).  }
\label{fig:rixs}
\end{figure}

Experimentally the magnetic RIXS signal is already already in the undoped cuprate La$_{2}$CuO$_4$ --in fact it is strongest in this case. The magnetic loss features are at energies well below the charge gap of this magnetic insulator, at energies where the charge response function $S({\bf q},\omega)$ vanishes, as well as the {\it longitudinal} spin one --which is in fact a higher order charge response function. This implies that we need take the ultra-short lifetime expansion~\cite{Brink05,Ament06} one step further.
By doing so I find the scattering amplitude for magnetic RIXS, expressed in terms of an intrinsic dynamic {\it four-spin correlation function} of the system that is probed. Moreover I derive selection rules which are related to the symmetry of the underlying spin Hamiltonian.  It turns out that the first allowed magnetic scattering process is a two magnon scattering one. As an example the formalism is used to calculate the indirect RIXS spectrum of Heisenberg antiferromagnets as function of transferred momentum $\bf q$. The scattering intensity vanishes at $(0,0)$ and at the antiferromagnetic wavevector ${\bf q}=(\pi,\pi)$. The computed spectra all agree with the experimental data. 
%This illustrates the point that the correlation functions and selection rules of magnetic RIXS are very different from neutron scattering --a technique that in essence measures a {\it two-spin} correlation function. 

{\it Series expansion of the scattering cross section.}
The probability for X-rays to be scattered from a solid state system can be enhanced by orders of magnitude when the energy of the incoming photons is in the vicinity of an electronic eigenmode of the system --i.e. in the vicinity of a resonance.  At a transition metal K-edge a $1s$ electron from the inner atomic core is excited into an empty $4p$ state, see Fig.~\ref{fig:rixs}. In transition metal systems the empty $4p$ states are far (10-20 eV) above the Fermi-level, so that the X-rays do not cause direct transitions of the $1s$ electron into the lowest $3d$-like  conduction bands of the system. Still this technique is sensitive to low energy excitations of the $d-$electrons because the Coulomb potential of $1s$ core-hole can couple to e.g. very low energy electron-hole excitations when the system is metallic. Since the charge excitations are caused by the core-hole, this scattering mechanism is sometimes referred to as {\it indirect} RIXS. 

In this Letter, however, we will consider insulating systems --in particular Mott insulators, where the only remaining low-energy degrees of freedom are the spin ones. So in order to determine the RIXS scattering amplitude we need to establish how in this case the core-hole couples to {\it magnetic} excitations. 
Our starting point is the Kramers-Heisenberg formula for the resonant scattering cross section~\cite{Kramers25,Note1}
\begin{eqnarray}
\left. \frac{ {\rm d}^2 \sigma} { {\rm d} \Omega {\rm d} \omega}  \right|_{res} & \propto &
\sum_{f}  | A_{fi} |^2 \ \delta (\omega-\omega_{fi}), \ \ {\rm with} \nonumber \\
A_{fi} &=& \omega_{\rm res} \sum_{ n} 
\frac{\langle f | \hat{D} | n \rangle \langle n | \hat{D} | i \rangle }
{\omega_{\rm in}-E_n -i \Gamma}. 
\label{eq:Kramers}
\end{eqnarray}
and $f$ and $i$ denote the final and initial state of the system, respectively. The sum is over $f$ is over all final states. The momentum and energy of the incoming/outgoing photons is ${\bf q}_{{\rm in}/{\rm out}}$ and $\omega^0_{{\rm in}/{\rm out}}$ and the loss energy $\omega= \omega^0_{\rm out}-\omega^0_{\rm in}$ is equal to the energy difference between the final and initial state $\omega_{fi}=E_f-E_i$.  In the following we will take the groundstate energy of our system as reference energy: $E_i=0$.
In the scattering amplitude $A_{fi}$  the resonant energy is $\omega_{\rm res}$, $n$ denotes the intermediate states and $\hat{D}$ is the dipole operator that describes the excitation from initial to intermediate state and the de-excitation from intermediate to final state. The dipole operator is given in more detail in for instance Ref.~\cite{Brink05}.
The energy of the incoming X-rays with respect to the resonant energy is $\omega_{\rm in}$ (this energy can thus
either be negative or positive: $\omega_{\rm in}= \omega^0_{\rm in}-\omega_{\rm res}$) and $E_n$ is the 
energy of intermediate state $|n\rangle$ with respect to the resonance energy. The last important detail is that the intermediate state is not a steady state. The highly energetic $1s$ core-hole quickly decays e.g. via Auger processes and the core-hole life-time is very short. This leads to a core-hole energy broadening $\Gamma$ which is proportional to the inverse core-hole life-time.

To calculate RIXS amplitudes, we proceed by formally expanding the scattering amplitude in a power series~\cite{Brink05}
\begin{eqnarray}
A_{fi} = 
\frac{ \omega_{\rm res} } {\Delta}  \sum_{l=1}^{\infty} \frac{1}{\Delta^l} 
\langle f | \hat{D} (H_{\rm int})^l \hat{D} | i \rangle
\label{eq:A_fi}
\end{eqnarray}
where we introduced $\Delta=\omega_{\rm in}-i\Gamma$ and the Hamiltonian in the intermediate state $H_{\rm int}$. For a further expansion of this scattering amplitude it is essential that we split up the intermediate state Hamiltonian into two parts: $H_{\rm int}= H_0 + H_1$, where $H_0$ is the Hamiltonian of the system without core-hole and $H_1$ the part of the Hamiltonian that is active in the presence of a core-hole. 

{\it Spin Hamiltonian with core-hole.}
We will calculate the resonant X-ray cross section in a Mott-Hubbard insulator at zero temperature. We assume that this system is described a single band Hubbard model at strong coupling and at half filling. In this case the electrons are localized and the only low energy degree of freedom is their spin. It is well known that in a Mott-Hubbard insulator the magnetic exchange integrals are determined by a virtual hopping process of electrons. We denote the hopping amplitudes of the valence electrons by $t_{ij}$ where $i$ and $j$ denote lattice sites with lattice vectors ${\bf R}_i$ and ${\bf R}_j$. The Coulomb interaction between electrons at the same site is $U$, so that in second order perturbation theory we have the exchange interaction $J_{ij}= 2 t_{ij}^2/U$ and the spin dynamics is governed by a Heisenberg spin Hamiltonian of the form
\begin{eqnarray}
H_0 =   \sum_{i,j} J_{ij}  {\bf S}_i {\bf S}_j = \sum_{\bf k} J_{\bf k} \ {\bf S_k} \cdot {\bf S_{-k}},
\label{eq:H0}
\end{eqnarray}
where $J_{\bf k}$ is the Fourier transform of $J_{ij}$. It is well known that for neutron scattering on such a spin system the {\it two-spin} correlation function $\sum_{\alpha}\int e^{-i \omega t} \langle S_{\bf q}^{\alpha}(0)  S_{-{\bf q}}^{\alpha}(t) \rangle dt$ is measured, where the sum $\alpha$ is over the three spin components. We will see shortly that magnetic RIXS measures a very different {\it four-spin} correlation function.
 
In the intermediate state a core-hole is present. We assume the core-hole potential $U_c$ to be local, i.e. as acting exclusively on those (valence) electrons that belong to the atom with the core-hole. When on site $m$ a core-hole is present, the exchange interactions that involve the spin on site $m$ becomes stronger, as the virtual intermediate state with two electrons on the site with the core-hole are lowered in energy by $U_c$. On the other hand the virtual state with two holes present on site $m$ is at $U+U_c$. Adding these two effects leads to the following exchange interaction between the spins on site $m$ and $j$ of
\begin{eqnarray}
J^c_{mj}=  2t_{mj}^2 \frac{U}{U^2-U_c^2}=  (1+\eta) J_{mj}
\end{eqnarray} 
and $\eta=\frac{U_c^2}{U^2-U_c^2}$.
From this we obtain $H_1$, part of the intermediate state Hamiltonian that is active in the presence of a core-hole
\begin{eqnarray}
H_1 = \eta \sum_{m,j} s_m s^{\tiny \dagger}_m J_{mj} {\bf S}_m \cdot {\bf S}_j,
\label{eq:H1}
\end{eqnarray}
where the operator $s_m$ creates a core-hole on site $m$. 

\begin{figure}
\begin{center}
\includegraphics[angle=-90,width=0.78\columnwidth]{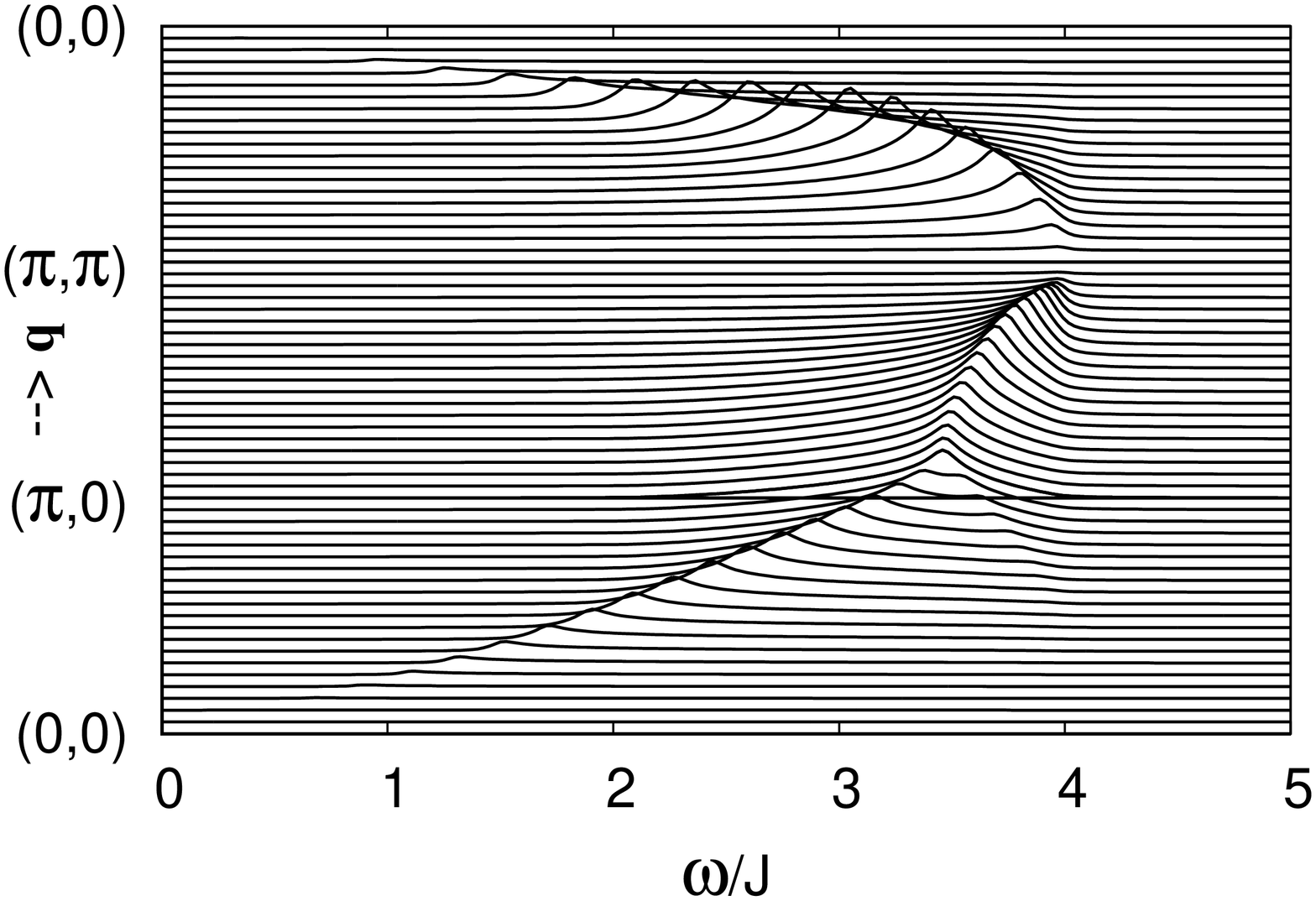}\\
%\vspace{-1cm}
\includegraphics[width=0.20\columnwidth,viewport=40 -170 180 180]{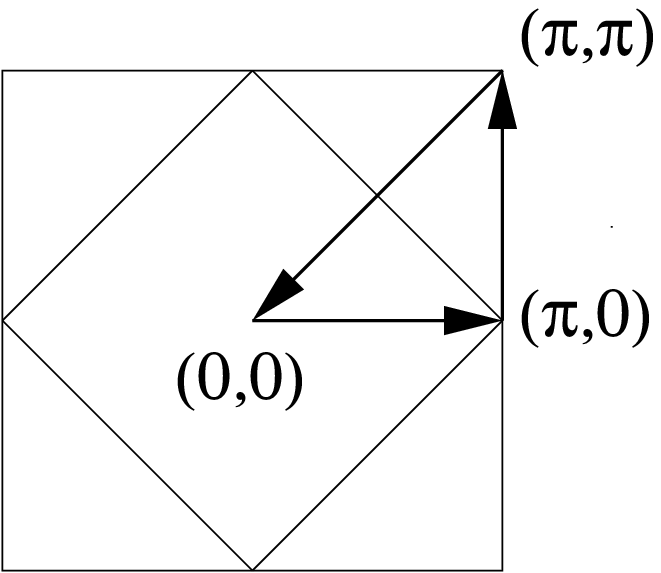}
\end{center}
\caption{Left: RIXS spectrum for a nearest neighbor Heisenberg antiferromagnet with exchange interaction $J$ as a function of transferred momentum $\bf q$ for a cut through the Brioullin zone that is show on the right.}
\label{fig:spectrum}
\end{figure}

{\it Spin-spin correlation function as measured in RIXS}.
In order to finally obtain the magnetic cross section, we need to evaluate the operator $(H_{\rm int})^l$ in equation (\ref{eq:A_fi}).  This is a non-trivial task. We first expand  $(H_{\rm int})^l$ in a series that contains the leading terms to the scattering cross section in lowest order in $\eta J/\Delta$. A conservative estimate gives, using for the copper K-edge $\Gamma \approx 1.5 \ eV$  and for the high temperature superconductors $J \approx 125 \ meV$ and $U_c/U \approx 0.85$ \cite{Hamalainen00,Benedetti01},  that at resonance $\eta J/\Delta \approx 0.22$. This makes it a suitable expansion parameter. Re-summing the leading order terms in the series gives in the end the magnetic scattering amplitude
\begin{eqnarray}
A_{fi}= {\omega_{\rm res} \over \Delta} {\eta \over \Delta-\omega} \langle f | \hat{O}_{\bf q} | i \rangle
\end{eqnarray}
where we find that the scattering operator $\hat{O}_{\bf q}$ to be
\begin{eqnarray}
\hat{O}_{\bf q} = \sum_{\bf k} J_{\bf k} \ {\bf S_{k-q}} \cdot {\bf S_{-k} }. 
\label{eq:Omagn}
\end{eqnarray}
so that the  magnetic correlation function that is measured in RIXS is proportional to the spin correlator $\int e^{-i \omega t}\langle \hat{O}_{\bf q}(0)  \hat{O}_{-{\bf q}}(t) \rangle dt$.

This expression is surprisingly simple and elegant. It shows that indeed momentum resolved indirect RIXS probes a momentum dependent four-spin correlation function. From expression~(\ref{eq:Omagn}) it is immediately clear why experimentally the magnetic RIXS intensity vanishes at zero transferred momentum, i.e. at ${\bf q} =0$. In that case the correlation function is nothing but the steady state Hamiltonian of the system ($\hat{O}_{{\bf q}=0} \propto H_0$).  Thus $|i\rangle$ and $|f\rangle$ are eigenstates of this magnetic scattering operator, which makes inelastic scattering impossible~\cite{Note2}.  This is in stark contrast with conventional two-magnon Raman scattering in the optical or UV range. That technique is also sensitive to a four-spin correlation function~\cite{Cottam72}, but a quite different one. This is obvious considering the fact conventional Raman scattering is restricted to ${\bf q}=0$ --precisely the momentum transfer where RIXS vanishes. Therefore also these two techniques offer complementary information on spin dynamics.
%Another consequence of form of the four-spin correlation function is that magnetic RIXS is absent for any fully saturated ferromagnet. The reason is again that in this case the groundstate is an eigenstate of the scattering operator $\hat{O}_{\bf q}$.

{\it Two-magnon scattering in antiferromagnets}.
From equation~(\ref{eq:Omagn}) immediately another selection rule follows. The projection of the total spin on the $z$-axis, $S^z_{tot}=\sum_i S^z_i$ commutes with both $H_0$ and $\hat{O}_{\bf q}$. Therefore $S^z_{tot}$ is conserved during the scattering process, which implies that creation of a {\it single} magnon by the core-hole is not possible. But the creation of two magnons (with opposite $z$-projections) is allowed and this is therefore the lowest order transversal spin scattering process that contributes to indirect RIXS. Also four-magnon scattering is in principle allowed, but of higher order and therefore smaller and not taken into account in the linear spinwave analysis that follows. Note that on the grounds of symmetry it is possible, in principle, to have magnetic scattering {\it without} creating any additional magnons in the scattering process. Physically this situation can only arise at finite temperature, when a magnon with momentum $\bf k$ that is present in the groundstate is scattered to ${\bf k}+{\bf q}$ by the core-hole. This implies that magnetic RIXS has an interesting temperature dependence --which is, however, beyond our present scope.

We now apply the theory above to two-dimensional bipartite $S=1/2$ antiferromagnets and determine the two-magnon RIXS spectrum as a function of transferred momentum, at zero temperature.
To this end the Hamiltonian $H_0$ and correlation function $\hat{O}_{\bf q}$ are bosonized within linear spinwave theory, where $
S^+_i \rightarrow a^{+}_i,  S^-_i \rightarrow a_i$ and $S^z_i \rightarrow {1 \over 2} -n_i$,
with boson creation/annihilation operators $a_i/a^+_i$ and number operator $n_i=a^+_ia_i$. After a Boguliobov transformation into the boson operators $\alpha_i/\alpha^+_i$ we have
$
\alpha^{+}_{\bf k}= u_{\bf k} a^{+}_{\bf k} + v_{\bf k} a_{\bf -k}
$
with
\begin{eqnarray}
u_{\bf k} = \sqrt{ {J_{{\bf k}=0} \over \epsilon_{\bf k}}+ {1 \over 2} },\ \
v_{\bf k} = {\rm sign[J_{\bf k}]} \sqrt{ {J_{{\bf k}=0} \over \epsilon_{\bf k}}- {1 \over 2} } 
\end{eqnarray}
and $\epsilon_{\bf k}=2 \sqrt{ J_{{\bf k}=0}^2 - J_{\bf k}^2 }$, then
the Hamiltonian reduces to
$
H_0^{LSW} =   \sum_{\bf k} \epsilon_{\bf k}  \alpha^+_{\bf k} \alpha_{\bf k}.
$
It is straightforward to show now that within linear spinwave theory the two magnon part of the magnetic scattering operator is
\begin{eqnarray}
\hat{O}^{LSW}_{\bf q}&=&   \sum_{{\bf k}>0} (J_{\bf k-q/{\rm 2}}+J_{\bf k+q/{\rm 2}}) 
(u_{\bf k-q/{\rm 2}}u_{\bf k+q/{\rm 2}} \nonumber \\ 
&+&v_{\bf k-q/{\rm 2}}v_{\bf k+q/{\rm 2}})-(J_{0}+J_{\bf q}) 
(u_{\bf k-q/{\rm 2}}v_{\bf k+q/{\rm 2}} \nonumber \\ 
&+&v_{\bf k-q/{\rm 2}}u_{\bf k+q/{\rm 2}}) \left( \alpha^+_{\bf k-q/{\rm 2}}\alpha^+_{\bf -k-q/{\rm 2}} + h.c. \right)
\label{eq:OLSW}
\end{eqnarray}
where $\bf q$ is the total momentum of the two magnon excitation. The resulting RIXS spectrum is shown in Fig.~\ref{fig:spectrum}, for a cut through the Brioullin zone indicated by the right-hand side of the figure. There are several remarkable features in the spectrum. 

\begin{figure}
\includegraphics[width=0.8\columnwidth]{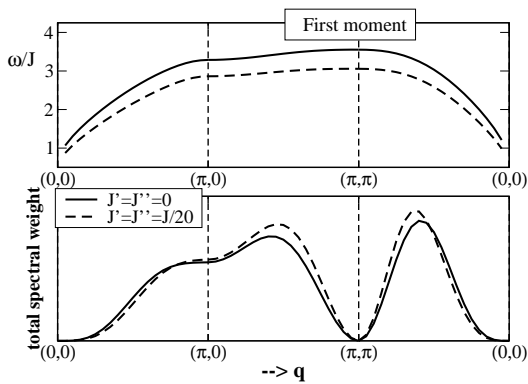} 
\caption{Top: first moment of the RIXS spectrum and bottom: total spectral weight for a nearest neighbor Heisenberg antiferromagnet (full line) and a Heisenberg model with in addition a second and third neighbor exchange ($J^{\prime}$ and $J^{\prime \prime}$, respectively) (dashed line). }
\label{fig:weight}
\end{figure}

First of all the spectral weight vanishes at ${\bf q} =(0,0)$ and ${\bf q} =(\pi,\pi)$. This in agreement with the experimental observations~\cite{Hill05}. From the scattering operator~(\ref{eq:OLSW}) it one sees that this selection rule is due to the antiferromagnetic ordering that occurs in the Heisenberg Hamiltonian. 
%In the scattering operator~(\ref{eq:OLSW}), the terms $J_{\bf k-q/{\rm 2}}+J_{\bf k+q/{\rm 2}}$ and  $J_{0}+J_{\bf q}$ are strictly zero at ${\bf q} =(\pi,\pi)$. 
It is actually easy to show that the RIXS intensity always vanishes at $(\pi,\pi)$ if this scattering vector is also a reciprocal lattice vector. This holds for instance also for a Heisenberg Hamiltonian with weak second and third neighbor exchange intereactions ($J^{\prime}$ and $J^{\prime \prime}$, respectively), which is illustrated by the calculation RIXS spectrum for an extended Heisenberg antiferromagnet with $J^{\prime}=J^{\prime \prime}=J/20$, shown in Fig.~\ref{fig:weight}. The longer range couplings transfer spectral weight to scattering vectors around $(\pi,\pi)$, but do not induce weight at precisely that wavevector.

The other remarkable feature feature of the magnetic RIXS spectrum is its strong dispersion. This is apparent from Fig.~\ref{fig:spectrum} and the upper pannel of Fig.~\ref{fig:weight}, which shows the first moment (average peak position) of the spectrum. The calculations for the nearest neighbor Heisenberg antiferromagnet (Fig.~\ref{fig:weight}) show that the magnetic scattering disperses from about $\omega \approx 0$ around $(0,0)$ and $(\pi,\pi)$ to $\omega \approx 4J$ at $(\pi,0)$ and $(\pi/2,\pi/2)$. Longer range couplings tend to reduce the first moment of the RIXS spectrum.
The observed dispersion has a two-fold origin. It is in part due to the $\bf q$-dependence of the two-magnon density of states (DOS),  combined with the scattering matrix elements that tend to pronounce the low energy tails of the two-magnon DOS. 
%In spite of being a two-magnon scattering process, the RIXS dispersion seems to  resemblance the single magnon dispersion.
%This is attributed to the form of the scattering operator $\hat{O}_{\bf q}$ --in particular that it contains on the magnetic couplings $J_{\bf k}$, which in this situation play the role of scattering matrix elements.

The consistency at ${\bf q}=(0,0)$ and ${\bf q}=(\pi,\pi)$ of the theoretical results and experimental data was already noted, but at other wave-vectors the agreement stands out even more. The data on La$_2$CuO$_4$ show for ${\bf q}=(\pi,0)$ a peak at around 500 $meV$, precisely where we find it on the basis of a nearest neighbor Heisenberg model with $J=146 \ meV$ --a value also found by the analysis of neutron scattering data~\cite{Coldea01}. Similar agreement is found at ${\bf q}=(0.6\pi,0)$ and ${\bf q}=(0.6\pi,0.6\pi)$. 
%Higher order exchange effects such as for instance the four-spin cyclic exchange, which are quite appreciable for this compound~\cite{Toader05}, will have to be taken into account for a more detailed analysis. 

{\it  Conclusions.}
We determined the momentum dependent four-spin correlation that is measured in magnetic RIXS. On the basis of this the magnetic RIXS spectrum was calculated for Heisenberg antiferromagnets with short and longer range couplings. We derive selection rules show that only scattering processes that involve at least two magnons are possible and that the scattering intensity vanishes at zero momentum transfer and at the antiferromagnetic lattice vector --which are observed in experiment. Moreover theory and experiment agree very well on the two-magnon peak position and spectral weight throughout the measured Brioullin zone.
These results show that RIXS is in principle a powerfull tool to obtain new information on spin dynamics --information that is complementary to what can be obtained by other techniques such as neutron, non-resonant X-ray or conventional two-magnon Raman scattering.

{\it Acknowledgments.}
I thank John P. Hill and Michel van Veenendaal for stimulating discussions and sharing unpublished data. I thank Fiona Forte and Luuk Ament for critically reading the manuscript.
%I am grateful to the Theory Institutes at Brookhaven National Laboratories and Argonne National Laboratories for their hospitality. 
This work is supported by FOM and the Dutch Science Foundation.


\begin{references}
\bibitem{Kotani01} For a review see: A. Kotani and S. Shin, Rev. Mod. Phys. {\bf 73}, 203 (2001).
\bibitem{Hasan00} M.Z. Hasan {\it et al.}, Science {\bf 288}, 1811 (2000). 
\bibitem{Hasan02} M.Z. Hasan  {\it et al.},  Phys. Rev. Lett {\bf 88} 177403 (2002).
\bibitem{Kim02} Y.J. Kim {\it et al.}, Phys. Rev. Lett. {\bf 89}, 177003 (2002).
\bibitem{Hill98} J.P. Hill {\it et al.}, Phys. Rev. Lett. {\bf 80}, 4967 (1998)
\bibitem{Hamalainen00}K. H\"am\"al\"ainen {\it et al.}, Phys. Rev. B {\bf 61}, 1836 (2000).
\bibitem{Inami03} T. Inami {\it et al.}, Pys. Rev. B {\bf 67}, 45108 (2003).
\bibitem{Isaacs96} E.D. Isaacs {\it et al.}, Phys. Rev. Lett. {\bf 76}, 4211 (1996).
\bibitem{Abbamonte99} P. Abbamonte {\it et al.}, Phys. Rev. Lett. {\bf 83}, 860 (1999).
\bibitem{Tsutsui03} K. Tsutsui, T. Tohyama and S. Maekawa, Phys. Rev. Lett. {\bf 91}, 117001 (2003). 
\bibitem{Doering04} G. D\"oring {\it et al.}, Phys. Rev. B {\bf 70}, 085115 (2004). 
\bibitem{Brink05} J. van den Brink and M. Veenendaal, Europhysics Letters, {\bf 73}, 121 (2006); cond-mat/0311446. 
\bibitem{Ament06} L. Ament, F. Forte and J. van den Brink, cond-mat/0609767 (2006).
\bibitem{Hill05} J.P. Hill, G. Blumberg, Y.-J. Kim, D. Ellis, S. Wakimoto, R.J. Birgeneau, S. Komiya, Y. Ando, D. Casa, T. Gog, unpublished.
\bibitem{Kramers25} H.A. Kramers and W. Heisenberg, Z. Phys. {\bf 31}, 681 (1925).
\bibitem{Note1} Note that where possible indices and constants are suppressed to avoid cluttering in the equations.
\bibitem{Benedetti01} P. Benedetti, J. van den Brink, E. Pavarini, A. Vigliante, and P. Wochner, Phys. Rev. B. {\bf 63}, 60408(R) (2001), R. Caciuffo, L. Paolosini, A. Solier, P. Ghigna, E. Pavarini, J. van den Brink and M. Altarelli, 
Phys. Rev. B {\bf 65}, 174425 (2002). 
\bibitem{Note2} This  holds for the expansion in lowest order of $\eta J/\Delta$; higher order terms might in principle give rise to some residual spin related scattering intensity at ${\bf q}=0$.
\bibitem{Cottam72} See, for example: M.G. Cottam, J. Phys. C {\bf 5}, 1461 (1972) and references therein.
\bibitem{Coldea01} R. Coldea {\it et al.}. Phys. Rev. Lett. {\bf 86}, 23, 5377  (2001).
%\bibitem{Toader05} A.M. Toader {\it et al.}, Phys. Rev. Lett. {\bf 94}, 197202 (2005).
\end{references}
\end{document}